\documentclass[a4paper,12pt]{nature2}
\usepackage{graphicx}
\usepackage{times}
\usepackage{color}
\usepackage{amsmath,amssymb,bm}
\usepackage{here}
\usepackage{multirow}
\usepackage{comment}
\usepackage{subfigure}
\usepackage{physics}
\usepackage{ulem}

\usepackage{xr}

\makeatletter
\newcommand*{\addFileDependency}[1]{
  \typeout{(#1)}
  \@addtofilelist{#1}
  \IfFileExists{#1}{}{\typeout{No file #1.}}
}
\makeatother

\title{Observation via spin Seebeck effect of macroscopic magnetic transport from emergent magnetic monopoles}
\date{}

\author{
Nan~Tang $^{1}$, 
Stephan Glamsch$^{1}$, 
Aisha~Aqeel$^{1}$, 
Ludwig Scheuchenpflug$^{1}$, 
Michael Schulze $^{3}$, 
Christoph Liebald $^{4}$,
Daniel Rytz $^{4}$,
Christo Guguschev$^{3}$,
Manfred Albrecht$^{1}$,
Philipp Gegenwart$^{1}$
}

\begin{document}

\maketitle

\begin{affiliations}
\item Center for Electronic Correlations and Magnetism, University of Augsburg, Augsburg 86159, Germany
\item Leibniz-Institut für Kristallzüchtung (IKZ), 12489 Berlin, Germany
\item EOT GmbH–Coherent, 55743 Idar-Oberstein, Germany

\end{affiliations}

\newcommand{\dto}{Dy${}_{2}$Ti${}_{2}$O${}_{7}$}
\newcommand{\one}{[1$\overline{1}$0]}
\newcommand{\gl}{${\textcolor{green}{{\maltese}}}$}
\newcommand{\VSSE}{$V_{\rm SSE}$}
\newcommand{\przro}{Pr${}_{2}$Zr${}_{2}$O${}_{7}$}

\begin{abstract}
Magnetic monopoles, elusive in high-energy physics, have been realised as emergent quasiparticles in solid-state systems, where their unique properties hold promise for novel spintronic applications. 
Magnetic monopoles have been invoked in diverse platforms, including skyrmion lattices, chiral magnets, soft ferromagnets, and aritifical nanomagnets. Yet, a demonstration of their role in magnetic transport has remained elusive.
Here, we report such an observation via the spin Seebeck effect in the bulk insulating pyrochlore oxide, spin ice \dto. 
By applying a thermal gradient perpendicular to a [111]-oriented magnetic field, we detect a transverse spin Seebeck voltage marked by a dominant peak at the onset of monopole proliferation, accompanied by a secondary feature and frequency-dependent behavior.
Our findings establish a direct link between monopole dynamics and magnetic transport in an insulating medium, establishing a new pathway for probing fractionalized excitations and advancing towards novel spintronic applications.
\end{abstract}





\newpage
\section*{Introduction}

Magnets with strong fluctuations can give rise to exotic fractionalized excitations, where the system’s quantum numbers split into smaller constituents, such as spin 1/2 spinons in one dimensional quantum spin chains \cite{Tennant_1995, Giamarchi_2003, Hirobe_2017}, Majorana fermions in two dimensional Kitaev quantum spin liquids \cite{Kitaev_2006, Nasu_2016, Banerjee_2017, Do_2017, Hentrich_2018}, emergent magnetic monopoles in three dimensional pyrochlore spin ices \cite{TSakakibara_2003, CCastelnovo_2008, Morris_2009, Fennell_2009, CCastelnovo_2012, Tang_2023}. Such excitations are of broad interest because they cannot be described within the conventional Landau paradigm of symmetry breaking and local order parameters. They may also exhibit long lifetimes and low dissipation\cite{Giamarchi_2003, Kish_2025}, making them attractive both for fundamental studies and for potential spintronic applications. 

However, detecting fractionalized excitations is challenging. Traditional probes such as inelastic neutron scattering couple to local spin-flip operators ($\Delta S=1$). In systems where the quasiparticles are fractionalized, these probes excite composite states rather than individual quasiparticles. As a result, experiments often reveal broad continua or indirect signatures, rather than sharp, well-defined dispersion relations. Moreover, when fractionalized excitations are charge-neutral, conventional charge transport measurements fail to detect them, while thermal transport, though sensitive, is easily masked by phonons and other `background'  contributions.

In this work, we employ the spin Seebeck effect (SSE) \cite{Uchida_2008, Bauer_2012, Uchida_2014} as a complementary probe to detect magnetic thermal transport involving fractionalized excitations in a frustrated magnet. In SSE, a temperature gradient applied to the material under investigation creates thermal excitations which carry spin angular momentum. This flow of spin angular momentum is then injected into an adjacent heavy metal layer (e.g., Pt) with strong spin-orbit coupling, where it is converted into a charge current via the inverse spin Hall effect, enabling electrical detection of the magnetic transport of the target material as a transverse voltage\cite{Saitoh_2006}. SSE thus provides a high-resolution, table-top method for probing magnetic transport involving fractionalized excitations. Unlike conventional neutron or thermal transport techniques, SSE directly converts the flow of spin angular momentum into an electrical signal, enabling sensitive detection of fractionalized excitations in quantum magnets. Although phonons and other degrees of freedom can also contribute, SSE is potentially less susceptible to such background signals than bulk thermal transport (as we will demonstrate below).

To date, SSE has been successfully applied to detect spin transport 
carried by magnons in a wide range of systems, including ferromagnets, antiferromagnets, paramagnets, and helimagnets\cite{Kikkawa_2023}. Later, it has been extended to a variety of quantum magnets, including Tomonaga--Luttinger liquids\cite{Hirobe_2017,wang2025}, spin-nematic states\cite{Hirobe_2019}, spin-Peierls systems\cite{Chen_2021}, magnon Bose–Einstein condensates\cite{Xing_2022}, as well as two-dimensional van der Waals magnets\cite{Wal_2024, He_2025}. Very recently, spin Seebeck studies have also emerged in the newly identified class of altermagnets\cite{Bai_2023, Liao_2024, Ruales_2025}. These advances highlight the versatility and sensitivity of SSE. 
Nevertheless, there have been no demonstrations of SSE as a tool for detecting fractionalized excitations beyond one dimension before now.

Here we present the first, to the best of our knowledge, magnetic transport signatures in SSE of three dimensional fractionalized quasiparticles, revealed through the detection of emergent magnetic monopoles in spin-ice systems. Spin ices are Ising magnets whose ground state is governed by ``2-in, 2-out'' spin correlations on each tetrahedron resulting in a macroscopically degenerate manifold. A single spin flip within this manifold creates a pair of topological defects-- ``3-in, 1-out'' and ``1-in, 3-out'' configurations, which correspond to a pair consisting of a magnetic monopole and an antimonopole\cite{CCastelnovo_2008}, as shown in Figure 1{\bf a}. These monopoles behave as point-like topological excitations that propagate in a deconfined fashion through the spin ice lattice, flipping tensionless ``Dirac string'' in the process, along which the ``2-in, 2-out'' rule is locally rearranged without being disturbed\cite{Jaubert_2009, Morris_2009}. This unique setting provides a clean and tunable environment to explore monopole dynamics and their  role in magnetic transport.

Previous studies have extensively investigated the dynamics of magnetic monopole motion in bulk spin ice materials represented by \dto{}, characterized by magnetic susceptibility\cite{Snyder_2003, Snyder_2004, KMatsuhira_2009, Matsuhira_2011, Yaraskavitch_2012, Bovo_2013, Kassner_2015, Billington_2025} which probes bulk magnetization relaxation, and flux-noise measurements\cite{Dusad_2019, Samarakoon_2022, Hsu_2024} which detects time-dependent magnetization/monopole-density fluctuations. Additionally, neutron scattering has revealed the associated Dirac strings and monopole correlations\cite{Morris_2009, Kadowaki_2009} by measuring spin-correlation functions. Together these studies demonstrate that monopoles in spin ice can move diffusively and give rise to extremely slow magnetic relaxation at low temperatures. Complementary $\mu$SR experiments, which sense local internal fields at muon sites, have also been employed to probe monopole charge currents and their interactions, under the heading of  magnetricity and magnetolyte physics. Indeed, transposing the idea of the Wien-effect from electrolyte physics\cite{Bramwell_2009} led to an understanding of the nonequilibrium nonlinear susceptibility\cite{Kaiser_2015}. Moreover, low-temperature thermal conductivity measurements\cite{Kolland_2013} presented evidence of monopole mobility via their impact on the heat current.

Recent theoretical work has further refined our understanding: Tomasello et al.\ \cite{Tomasello_2019} showed that spins flip at two very different rates depending on their local environment. Building on this, Hall{{e}}n et al.\ \cite{Hallen_2022} found an initial subdiffusive motion of the monpoles on a dynamical fractal, followed by slower, conventional diffusion at large distances. This goes beyond the earlier picture of independent monopoles undergoing simple Brownian motion\cite{Ryzhkin_2005,Bovo_2013, Klyuev_2017}. Furthermore, in artificial spin-ice systems, monopole dynamics have been imaged directly by advanced microscopy techniques\cite{Farhan_2019, Arava_2020}, while magnetic force microscopy has uncovered static monopole defects and magnetic charge currents\cite{Ladak_2010, May_2021}.

Here, we report the observation of a thermally-driven fractionalized spin excitations in the spin ice \dto, detected via SSE measurements. As illustrated in Fig.~1{\bf b}, the device consists of a single crystal of \dto{} with a thin-film Pt Hall bar ($800 \times 50 \times 20~\mu$m) deposited on its surface. The Pt strip serves two purposes: when an ac current $J$ is applied, it locally heats the sample surface, thereby generating a temperature gradient across the sample; simultaneously, it acts as a detector, measuring the electric voltage (i.e.,  the spin Seebeck voltage $V_{\rm T}$ across the transverse Pt contacts) converted from the spin flow in the target material, observed as the second harmonic response to the applied current.
The temperature gradient is applied along the \one{} direction, which aligns with the $\alpha$ chains—two of the tetrahedral sublattices—allowing efficient spin propagation. Meanwhile, the magnetic field is applied in-plane along the [111] direction, known for its characteristic metamagnetic transition associated with monopole proliferation. This geometry is specifically chosen to efficiently probe the magnetic transport mediated by magnetic monopoles under thermally driven conditions.
For further experimental details, see the Methods section. These results provide direct evidence of magnetic transport carried by emergent magnetic monopoles in an insulating medium, highlighting a macroscopic manifestation of monopole dynamics.

\section*{Experimental results}
Since the spin Seebeck effect is closely linked to the magnetization behavior of the system, we have measured the magnetic field ($B$) dependence of the magnetization ($M$) and magnetic susceptibility ($dM/dB$) at 1.4 K for $B \parallel [111]$ in Dy$_2$Ti$_2$O$_7$, as shown in Fig. 2{\bf a}. The magnetization exhibits a rapid initial increase up to 0.3 T, followed by a shoulder feature corresponding to the kagome-ice plateau. In this low-field regime, the external magnetic field quickly polarizes one of the four spins in each tetrahedron, while the remaining three spins form a two-dimensional kagome-ice network with “1(2)-in, 2(1)-out” configurations. This reflects a partial lifting of the degeneracy of the ``2-in, 2-out'' spin-ice ground state, reducing it from sixfold to threefold. At higher fields, the system undergoes a metamagnetic transition into the fully polarized “1(3)-in, 3(1)-out” state, signaling the creation and proliferation of magnetic monopoles — often referred to as monopole condensation. The susceptibility $dM/dB$ clearly highlights this transition, with a pronounced peak at $B = 1.1$ T. This $B$-dependent behavior is well established in previous studies of classical spin-ice systems\cite{Matsuhira_2002, TSakakibara_2003}. The inset shows $dM/dB$ as a function of $B$ at selected temperatures. The peak associated with the metamagnetic transition diminishes with increasing temperature and disappears above 4 K.

Next, we present the main experimental result: the magnetic field ($B$) dependence of the spin Seebeck voltage ($V_{\rm SSE}$) in Pt/\dto{} devices, as shown in Fig.~2\textbf{b}. An AC current of 0.44~mA at 73~Hz is applied to the Pt strip, locally heating the sample and creating a thermal gradient. The hot side reaches approximately 20~K (see Extended Data Fig.~1), while the cold side is maintained using the He$^{3}$ cooling option in the PPMS. The sample is mounted directly onto a copper holder to ensure good thermal contact.

Weak spin Seebeck signals emerge around 15 K, as evidenced by the odd-in-$B$ response. This is consistent with the expected relation $V_{\rm SSE} \propto \mathbf{J}_S \times \mathbf{M}$, where $\mathbf{J}_S$ denotes the spin current. Down to 7.8 K, spin Seebeck voltage \VSSE{} scales with the magnetization (Extended Data Fig.~2), characteristic of a paramagnet at high-temperatures\cite{wang2025}. At $T_{\rm cold} = 4$ K, however, the system enters the spin-ice regime: a distinct peak appears near 0.3 T, marking the onset of the kagome-ice plateau. Notably, the amplitude of this peak is relatively insensitive to temperature. This peak likely originates from spin-flip processes within the ``2-in, 2-out'' manifold, which attempt to reorganize the various ``2-in, 2-out'' configurations, believed to be the dominant heat carriers. Specifically, at zero field, the spin ice ground state is sixfold degenerate. When a magnetic field is applied along the [111] direction, this degeneracy is partially lifted. Around 0.3 T, the system enters the kagome ice phase, characterized by a reduced, threefold-degenerate ground state. The observed peak in $V_{\rm SSE}$ reflects the entropy reduction associated with the rearrangement of these constrained Ising spins. While dilute thermal monopoles may still be present at these temperatures, their contribution is expected to be minor, as this peak remains temperature independent. This robustness suggests that the signal originates primarily from field-driven configurational changes rather than thermally activated quasiparticle transport. 

At $T_{\rm cold} = 2.4$~K, a second peak appears near 1.1~T, corresponding to the metamagnetic transition associated with monopole proliferation. This peak becomes more pronounced at lower temperatures, in agreement with the magnetic susceptibility $dM/dB$ shown in the inset of Fig.~2\textbf{a}. 
We propose that this feature is a signature of dynamical magnetic monopoles, which have a density that onsets exponentially as the field strength $B$ approaches the critical point $B_c$ \cite{Ryzhkin_2005}. 
Once the system enters the fully polarized phase beyond the transition, the monopole density saturates, leading to a suppression of their motion. This results in a pronounced peak in \VSSE{} at $B_{\rm c}$.
The monopole mobility generically scales with inverse temperature $\mu\sim1/T$ \cite{castelnovo2011,mostame2014,Hirobe_2017}, leading to enhanced transport as temperature is decreased.
This trend is consistent with our observation of \VSSE{} in Fig. 2\textbf{b}.
Further confirmation of the thermal origin of these peaks is that their amplitude scales with $I^2$ (Extended Data Fig.~3). 
Interestingly, the full width at half maximum (FWHM) of the peak in $V_{\rm SSE}$ is only about 60\% that of $dM/dB$. This narrower linewidth suggests that $V_{\rm SSE}$ probes a distinct subset of excitations -- specifically, mobile monopoles -- rather than reflecting all contributions to the static magnetization.
As a control, Fig.~2\textbf{c} shows the $V_{\rm SSE}$ measured on a non-magnetic Pt/SiO$_{\rm x}$ device under identical conditions. No spin Seebeck signal is detected, confirming that the observed effects originate from spin excitations in the \dto{} substrate.

Here, we examine the relationship between the \VSSE{} and the thermal conductivity\cite{Kolland_2013} ($\kappa$), both measured in the same geometry, with the magnetic field $B \parallel [111]$ and heat propagating along the \one{} direction (Extended Data Fig.~4). Importantly, while both are influenced by monopole dynamics under a thermal gradient, the probing mechanism is different: $\kappa$ measures bulk heat flow, whereas \VSSE{} selectively detects the magnetic transport by converting it to a charge current via the inverse spin Hall effect. The thermal conductivity $\kappa$ exhibits a drop between 0 and 0.3 T, followed by a kink near the metamagnetic critical field at approximately 1 T at 1 K, and then continues to decrease gradually up to 7 T. The characteristic field values of 0.3 T and 1 T are also observed in \VSSE. However, a key difference lies in the high-field behavior: while $\kappa$ shows a continuous field dependence, \VSSE{} returns to a nearly zero baseline beyond the peak structure. G. Kolland et al. demonstrated that the thermal conductivity of \dto{} above 1.5 T is dominated by a magnetic-field-dependent phononic background, as evidenced by similar high-field behavior in half-doped Dy$_{0.5}$Y$_{0.5}$Ti$_{2}$O$_{7}$. In contrast, \VSSE{} here do not pick up such phononic contributions, underscoring its sensitivity to purely magnetic transport in \dto.

To further investigate the peaks observed in \VSSE, we performed a frequency sweep at selected magnetic fields, shown in Fig.3. Remarkably, a broad peak centered around 0.5 $\sim$ 1 kHz was observed at $B = \pm 1.1$ T, corresponding to the metamagnetic anomaly.
The interfacial conversion between the spin dynamics in \dto{} and the charge current in Pt is expected to be effectively instantaneous within our measurement frequency range, as ultrafast studies on Pt/YIG systems have demonstrated sub-picosecond to picosecond timescales for buildup of spin Seebeck signal\cite{Seifert_2018,Kimling_2017}. Therefore, any kHz dispersion is unlikely to arise from interfacial processes and should instead reflect intrinsic dynamics within the spin-ice state. Near the critical field, the monopole density increases, but their motion remains slow due to local configurational constraints imposed by the spin-ice rules, resulting in subdiffusive motion with inherently long timescales\cite{Hallen_2022, Dusad_2019, Samarakoon_2022}. The broad SSE peak at $f= 1$ kHz reflects the system's characteristic response to a 2 kHz thermal excitation, as the SSE is detected at the second harmonic (2$f$) of the applied AC current. This corresponds to a characteristic relaxation time of $\tau \sim 1/2\pi 2f \sim 0.08 - 0.16$ ms. This timescale is consistent with the intrinsically slow monopole dynamics in \dto{}, as previously revealed by various measurements\cite{Snyder_2004, Matsuhira_2011, Yaraskavitch_2012, Kassner_2015, Eyvazov_2018, Dusad_2019, Samarakoon_2022, Hsu_2024}. The latest study of flux noise measurement\cite{Hsu_2024} shows a dichotomy between a fast reconfigurational process ($\sim$ 90 $\mu$s) and a slower monopole-driven mechanism (1 $\sim$ 10 ms). 
Our SSE derived time scale signal (0.08 $\sim$ 0.16 ms) falls between these regimes.

This frequency dependence stands in sharp contrast to that of conventional magnetic insulators such as YIG, where the SSE typically exhibits a low-pass behavior: the signal remains relatively constant across low frequencies and rolls off in the MHz range, reflecting magnon–phonon relaxation times of a few hundred picosecond in thin films\cite{Schreier_2016}. In \dto{}, however, the appearance of a peak points to a characteristic relaxation channel intrinsic to the spin-ice state. While the SSE has been widely studied in conventional magnetic insulators, its manifestation in \dto{} reveals new dynamical aspects of spin transport, shaped by frustration and emergent monopole excitations.


\section*{Summary and Perspectives}
In summary, we have observed macroscopic magnetic transport involving emergent magnetic monopoles in bulk pyrochlore spin ice \dto{} using spin Seebeck measurements. To our knowledge, this is the first demonstration of three-dimensional fractionalized excitations detected via the spin Seebeck effect (SSE). Our results highlight the power of SSE: it couples selectively to magnetic transport and thus serves as a complementary probe of exotic excitations in quantum matter. Because emergent magnetic monopoles can propagate without incurring additional energy costs, they present a promising platform for using fractionalized excitations as carriers of spin information in future spintronic applications.

\begin{addendum}
\item We thank R. Moessner and J. Wilsher for substantial discussions, and T. Sakakibara, A. Herrnberger, M. Althammer, H. Huebl, K. Uchida, and S. Wu for their helpful insights. Work at Augsburg is supported by the Alexander von Humboldt foundation and the Deutsche Forschungsgemeinschaft (DFG, German Research Foundation) through project —TRR 360-492547816 (B5) and DFG project No. 492421737. A.A. acknowledges the support by DFG Emmy Noether Programme. We gratefully acknowledge the funding for the single crystal growth development of \dto{} within the cooperation project “Pyrochlore crystals for compact optical isolators” provided by the Federal Ministry for Economic Affairs and Energy (former BMWi, now BMWE) due to an enactment of the German Bundestag under Grant No. ZF4728901RE9 and ZF4730501RE9 within the Central Innovation Program for SMEs (ZIM).

\item[Contributions] 
N.T. and P.G. conceived the project, N.T. designed and organized the experiments. M.S. led the single-crystal growth with support from C.L., D.R., and C.G.. N.T. oriented the samples and polished the surfaces. N.T., S.G., and L.S. performed Atomic Force Microscopy (AFM) imaging to evaluate surface roughness. S.G. carried out optical lithography and Pt Hall-bar sputtering. N.T. conducted magnetization measurements, and N.T., A.A., and L.S. performed spin Seebeck measurements. N.T. analyzed the data, prepared the figures, and wrote the manuscript with input from all authors. 

\item[Competing Interests] The authors declare that they have no competing interests.
\item[Corresponding authors]
Correspondence and requests for materials should be addressed to N. Tang. (email:nan.tang@uni-a.de).
\end{addendum}

\newpage
\begin{figure}[ht]
 \begin{center}
 \includegraphics[keepaspectratio, scale=0.7]{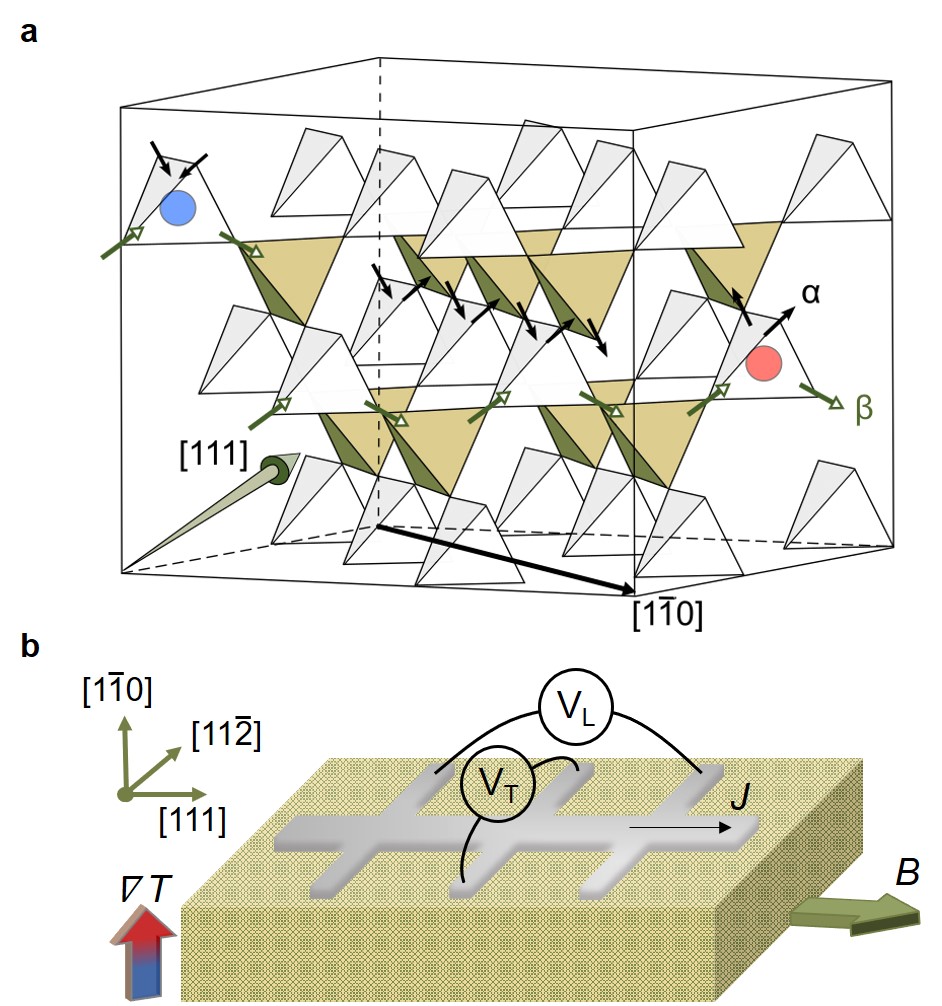}
 \end{center}
\end{figure}
 \normalsize{{\bf Figure 1 $|$ Schematic of the experimental geometry of the pyrochlore oxide \dto.}
 {\bf a}, Illustration of the pyrochlore lattice composed of corner-sharing tetrahedra, with crystallographic directions [111] and \one{} indicated. The magnetic moments (Ising spins) are constrained to lie along the local [111] axes of each Dy$^{3+}$ ion, forming a short-range spin-ice configuration characterized by the ``2-in, 2-out'' ice rule. The \one{} direction corresponds to the $\alpha$ chains (black arrows), while the perpendicular [110] direction corresponds to the $\beta$ chains (green arrows with hollow heads). An excited spin configuration of either ``1-in, 3-out'' or ``3-in, 1-out'' is also shown, representing positive and negative magnetic monopole charges, illustrated as blue and red spheres, respectively. 
 {\bf b}, Schematic of the measurement configuration for spin Seebeck effect (SSE) measurements. The device consists of a bulk \dto{} single crystal (bottom) with a 4-nm-thick Pt Hall bar deposited on top. An ac current is applied along the Pt strip, oriented along the [111] direction, generating a temperature gradient $\nabla T$ in the perpendicular \one{} direction. With a magnetic field $B$ applied along [111], the transverse voltage $V_{\rm T}$ (i.e., the spin Seebeck voltage \VSSE{}), measured as the second harmonic response of the applied current, is detected perpendicular to both the field and the thermal gradient, satisfying the orthogonality condition required for SSE. A longitudinal voltage, $V_{\rm L}$, is also measured along the current direction for thermometry purpose.}
\label{Fig:Fig1}

\newpage
\begin{figure}[ht]
 \begin{center}
 \includegraphics[keepaspectratio, scale=0.55]{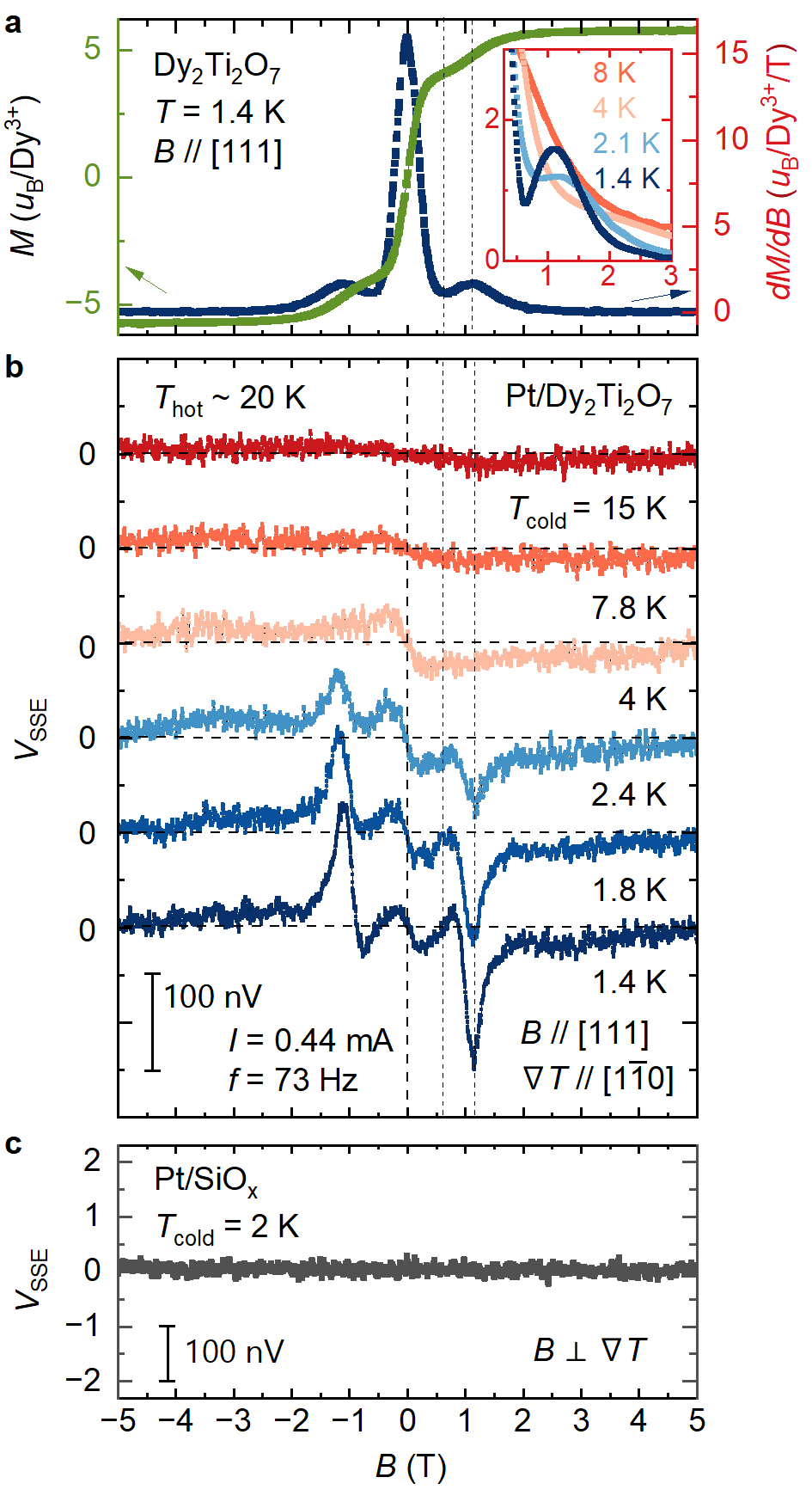}
 \end{center}
\end{figure}
 \normalsize{{\bf Figure 2 $|$ Magnetic field $B$ dependence of Magnetization $M$ and spin Seebeck voltage $V_{\rm SSE}$ in \dto.}
 {\bf a}, $M$ (left axis) measured at 1.4 K as a function of $B$ applied along the [111] direction. The corresponding magnetic susceptibility $dM/dB$ is shown on the right axis. Inset: $dM/dB$ curves at selected temperatures.  
 {\bf b}, $B$ dependence of the spin Seebeck voltage $V_{\rm SSE}$ of Pt/\dto{} device at selected temperatures. One major grid division corresponds to 100 nV.
 {\bf c}, $B$ dependence of $V_{\rm SSE}$ measured on a Pt/SiO$_{\rm x}$ device as a non-magnetic reference under identical conditions.
 The two finely dashed lines indicate characteristic magnetic field positions at 0.6 T and 1.1 T. The coarser dashed lines mark the zero points of both the $x$- and $y$-axes.
}
\label{Fig:Fig2}

\newpage
\begin{figure}[ht]
 \begin{center}
 \includegraphics[keepaspectratio, scale=0.4]{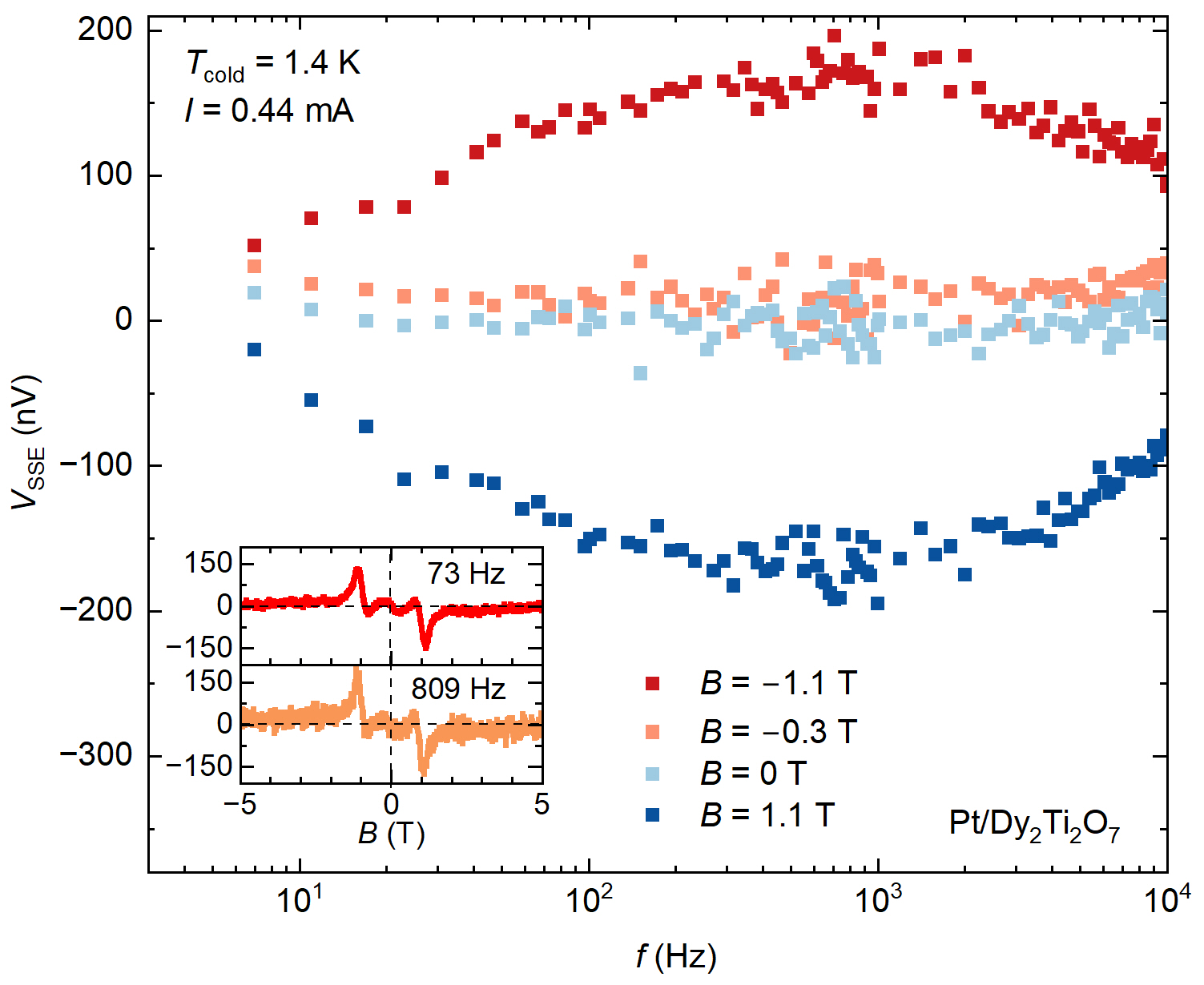}
 \end{center}
\end{figure}
 \normalsize{{\bf Figure 3 $|$ Frequency $f$ dependence of the characteristic peaks in $V_{\rm SSE}$.} $f$ scans of $V_{\rm SSE}$ are shown at selected magnetic fields where characteristic peaks are observed. The $x$-axis is plotted on a logarithmic scale. The inset shows the $B$ dependence of $V_{\rm SSE}$ at selected frequencies for consistency checks.
}
\label{Fig:FigKappa2}

\newpage
\bibliography{DTObib.bib}
\bibliographystyle{naturemag}

\clearpage
\begin{methods}
\section*{Sample preparation} 
A single crystal with the dimensions of 6 mm diameter and 44 mm length was grown in a Crystal Systems Corporation (CSC) optical floating zone furnace. Small rectangular pieces are cut from the single crystal rod. The crystal orientations were checked by a backscattering Laue X-ray diffractometer (Photonic Science) prior to the cutting. The sample is cut using a wire saw (DIDRAS) and subsequently polished with a polishing machine (ECOMET 30, BUEHLER), employing a fine polishing cloth (CHEMOMET, BUEHLER) and a non-crystallizing colloidal silica polishing suspension (MasterMet, BUEHLER). The sample surface roughness is recorded using an Dimension Icon AFM (Atomic Force Microscopy) from Bruker. See Extended Data Fig.~5 for a representative AFM picture. The 4 nm Pt hall bar was fabricated by optical lithography followed by magnetron sputter deposition. See Supplementary Information for more details regarding single crystal growth, surface cleaning and Pt dimensions and deposition process.

\section*{Control experiment on the non magnetic sample}
A Pt Hall bar of identical size and thickness was sputtered onto commercially purchased (111) Si substrates (0.7 mm thick) with a thermally grown amorphous SiO$_{\rm x}$ surface layer exhibiting an root-mean-square (RMS) surface roughness of around 700 pm, following the same procedure described above. No additional surface cleaning or surface treatment was applied.

\section*{Magnetization measurements}
Magnetization measurements of \dto{} were performed using a superconducting quantum interference device (SQUID) magnetometer in a Magnetic Property Measurement System (MPMS; Quantum Design), equipped with a He$^3$ insert that allows for a base temperature of 0.4 K. The magnetization of the Pt/SiO$_{\rm x}$ device was also measured prior to the spin Seebeck measurements shown in Fig. 2\textbf{c} in the main text, and no evidence of magnetic contamination was observed.

\section*{Spin Seebeck measurements}
Spin Seebeck voltages were measured in a Physical Property Measurement System (Dynacool; Quantum Design) equipped with a He$^3$ option, using standard low-frequency lock-in amplifiers (MFLI, Zurich Instruments). Two lock-in amplifiers were operated simultaneously: one recorded the longitudinal voltage, while the other measured the transverse voltage up to the second harmonic. The spin Seebeck signal was extracted from the second harmonic component with a –90° phase shift relative to the applied ac current. The method is well established in Ref. \citeonline{Vlietstra_2014}. The temperature gradient was generated via current heating (Fig. 1\textbf{b}, main text). The excitation current was supplied by an external Keithley 6221 AC/DC current source. We adopt the sign conventions of Ref. \citeonline{Schreier_2015} for our measurements.

\begin{addendum}
\item[Data Availability]
Source data that support the findings of this study are provided with this paper. 
\end{addendum}
\end{methods}

\clearpage

\newpage
\begin{figure}[ht]
 \begin{center}
  \includegraphics [keepaspectratio,scale=0.5] {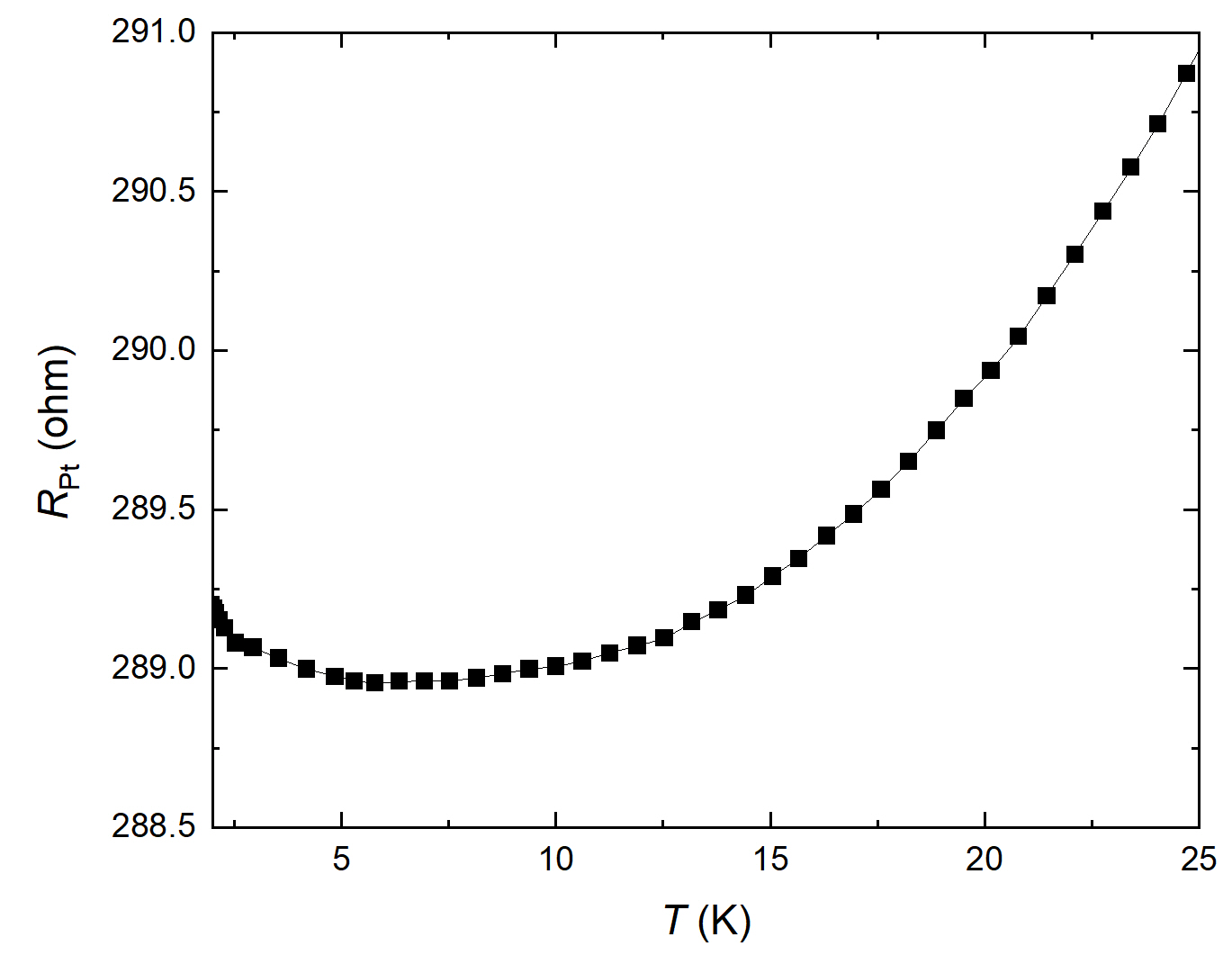}
\end{center}
\end{figure}
\normalsize{{\bf Extended Data Figure 1 $|$ Pt strip as a thermometer.} 
The temperature dependence of the resistance of the Pt strip ($R_{\rm Pt}$) acts as a rough thermometer for the device. Under heating with $I =0.44$ mA, the Pt strip shows a resistance of 289.8 ohm, which corresponds to a temperature of $\sim 20$ K. This estimate is consistent with the \VSSE{} data in Fig. 2\textbf{b}, where small SSE signals persist up to 15 K, indicating that the hot side is indeed above 15 K.}
\label{Fig:FigS2}

\newpage
\begin{figure}[ht]
 \begin{center}
  \includegraphics [keepaspectratio,scale=0.4] {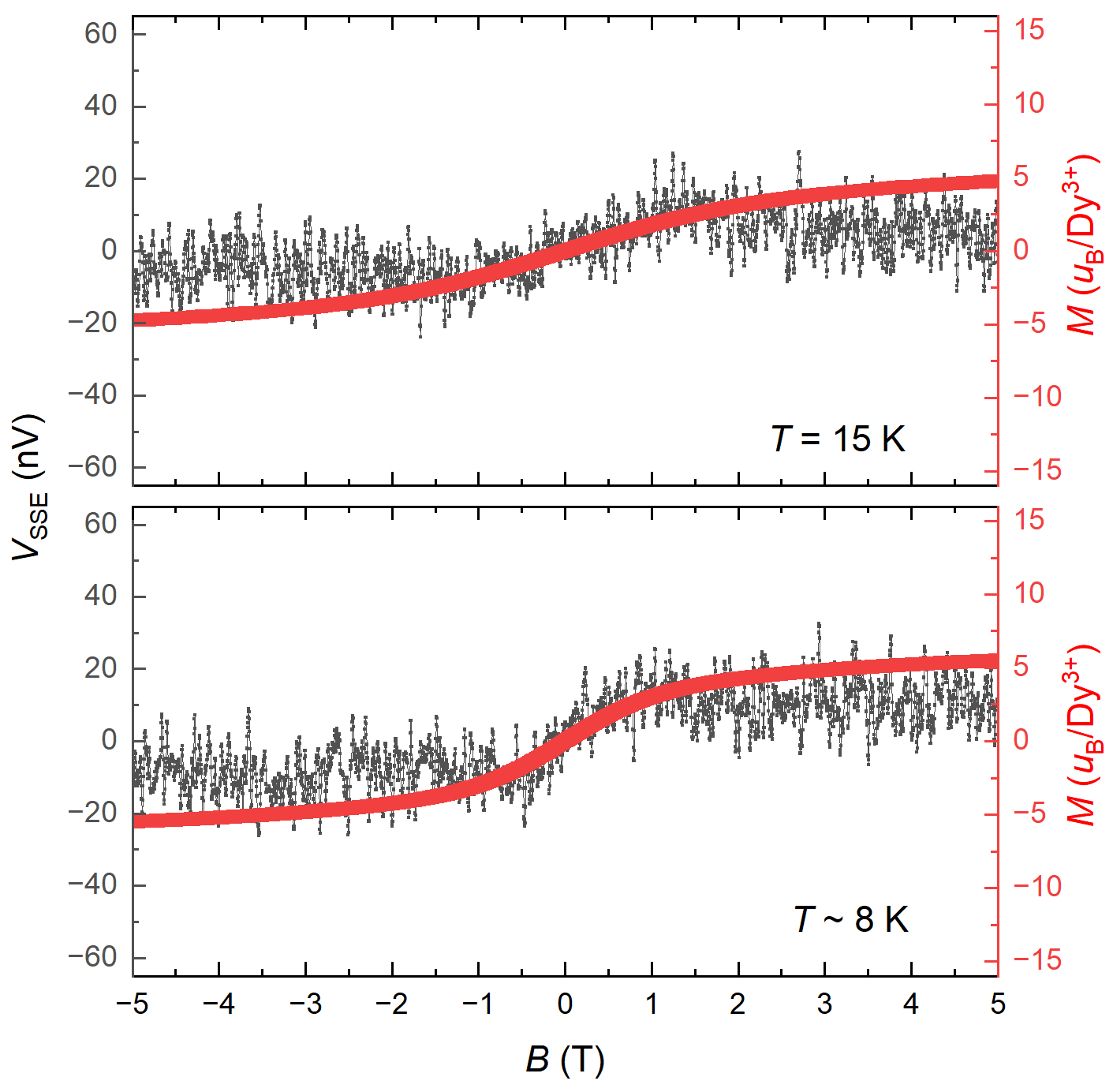}
\end{center}
\end{figure}
 \normalsize{{\bf Extended Data Figure 2 $|$  Comparison of spin Seebeck voltage (\VSSE{}) and magnetization $(M)$ at elevated temperatures.} The left axis shows the \VSSE{} signal, while the right axis shows $M$, at the high temperatures.} 
 \label{Fig:FigS4}

\newpage
\begin{figure}[ht]
 \begin{center}
  \includegraphics [keepaspectratio,scale=0.5] {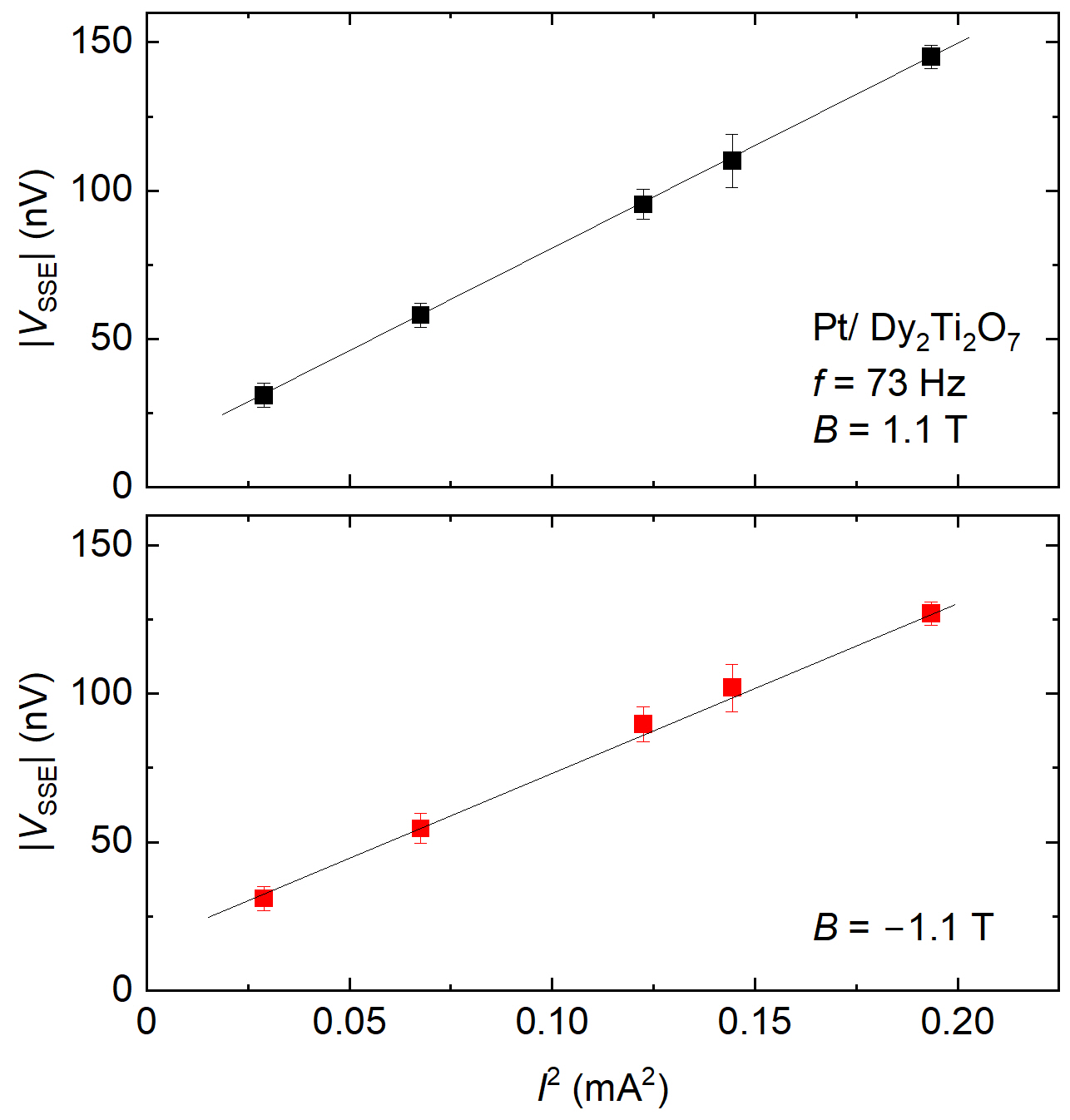}
\end{center}
\end{figure}
 \normalsize{{\bf Extended Data Figure 3 $|$ Quadratic current ($I^2$) dependence of the metamagnetic peak in $|$\VSSE{}$|$.} The metamagnetic peak at both $B = \pm 1.1$ T in $|$\VSSE{}$|$ follows a quadratic dependence on $I$, indicative of its thermal origin. Error bars are visually estimated from the noise fluctuations in the data. Black lines are provided as a guide to the eye.}
 \label{Fig:FigS3}

\newpage
\begin{figure}[ht]
 \begin{center}
  \includegraphics [keepaspectratio,scale=0.5] {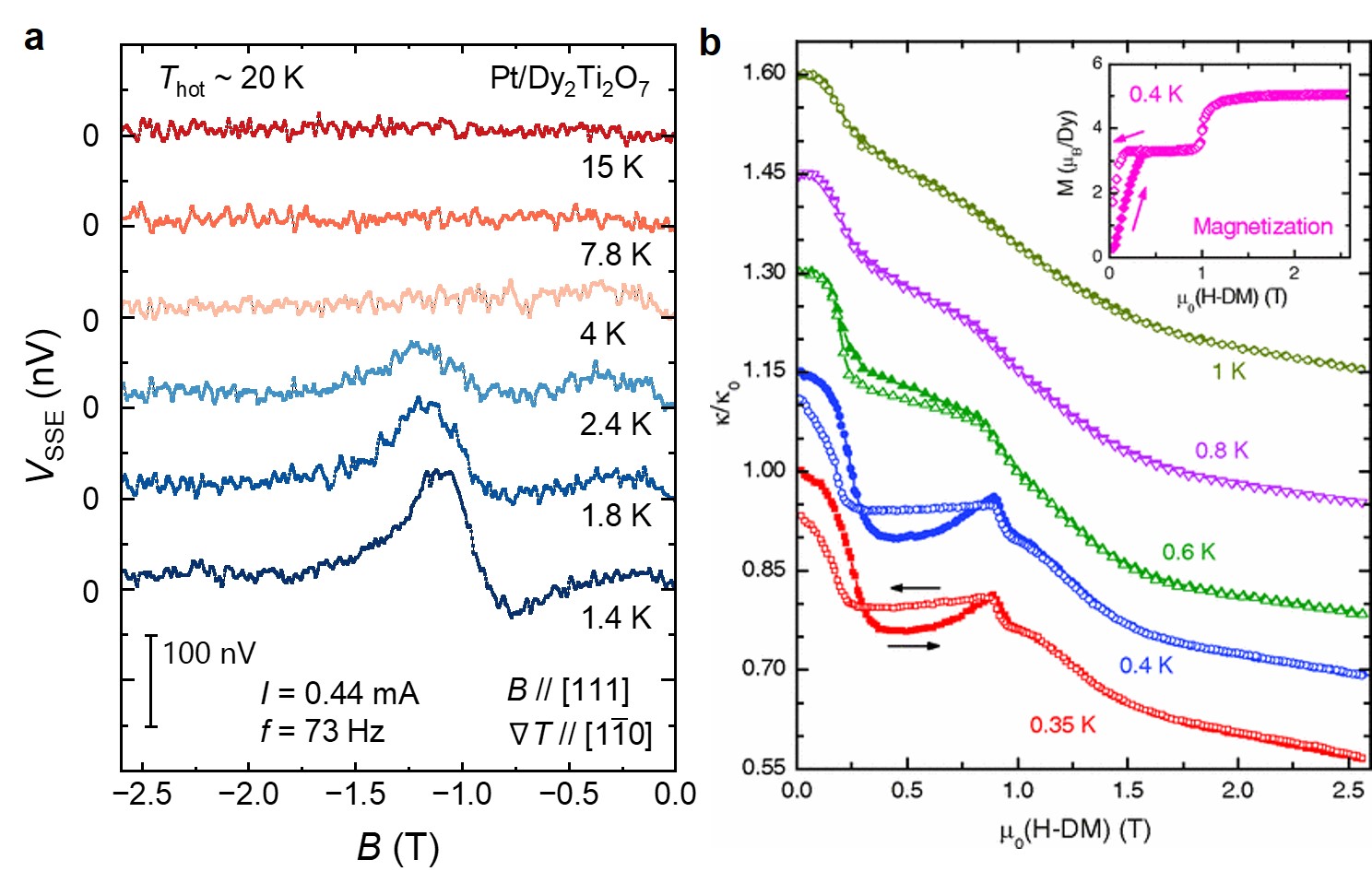}
\end{center}
\end{figure}
 \normalsize{{\bf Extended Data Figure 4 $|$ Comparison of thermal conductivity ($\kappa/\kappa_0$) and spin Seebeck voltage (\VSSE) in \dto. } Field ($B$) dependence of {\bf a}, \VSSE{} and {\bf b}, normalized thermal conductivity $\kappa /\kappa_0$, where $\kappa_0$ is zero-field value, under the same measurement geometry where $ B \parallel [111]$ and $\nabla T \parallel [1\overline{1}0]$. Panel {\bf b} is directly adapted from G. Kolland et al., Phys. Rev. B 88, 054406 (2013). Copyright (2013) by the American Physical Society. Reproduced with permission. DOI: 10.1103/PhysRevB.88.054406}
 \label{Fig:FigS5}

\newpage
\begin{figure}[ht]
 \begin{center}
  \includegraphics [keepaspectratio,scale=0.5] {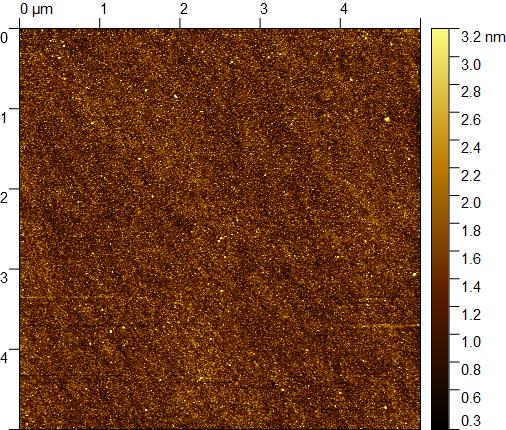}
\end{center}
\end{figure}
 \normalsize{
 {\bf Extended Data Figure 5 $|$ AFM image of \dto{} surface before lithography and Pt deposition.} 
 The root-mean-square (RMS) surface roughness $S_{\rm q}$ is $\sim$ \(430~\mathrm{pm}\).
 }
 \label{Fig:FigS1}

\end{document}